\documentclass[letterpaper,12pt]{article}
\usepackage{osajnl} 

\begin{document}

\title{A millimeter-wave antireflection coating for cryogenic silicon lenses}

\author{Judy Lau, Joseph Fowler, Tobias Marriage, and Lyman Page}
\affiliation{Princeton University}
\address{Department of Physics, Jadwin Hall\\
Princeton, NJ 08544, USA}

\author{Jon Leong}
\affiliation{Case Western Reserve University}
\address{Department of Physics\\
Rockefeller Hall 10900 Euclid Avenue\\
Cleveland, OH  44106, USA}

\author{Edward Wishnow}
\affiliation{Lawrence Livermore National Laboratory}
\address{7000 East Avenue\\
Livermore, CA 94551, USA}

\author{Ross Henry and Ed Wollack}
\affiliation{NASA Goddard Space Flight Center}
\address{Observational Cosmology Laboratory\\
Greenbelt, MD 20771, USA}

\author{Mark Halpern, Danica Marsden, and Gaelen Marsden}
\affiliation{University of British Columbia}
\address{Department of Physics and Astronomy\\
Vancouver, BC, Canada V6T-1Z1}

\begin{abstract} 
We have developed and tested an antireflection (AR) coating method for
silicon lenses at cryogenic temperatures and millimeter wavelengths.
Our particular application is a measurement of the cosmic microwave
background. The coating consists of machined pieces of Cirlex glued to
the silicon. The measured reflection from an AR coated flat piece is
less than 1.5\% at the design wavelength. The coating has been applied
to flats and lenses and has survived multiple thermal cycles from 300
to 4\,K. We present the manufacturing method, the material properties,
the tests performed, and estimates of the loss that can be achieved in
practical lenses.
\end{abstract}

\section{Introduction}
Silicon is a promising lens material for millimeter wavelength
observations because it can be machined; it has a high index of
refraction, which is optically advantageous; and it has a high thermal
conductivity, allowing for straightforward cooling of the lenses to
cryogenic temperatures. However, the high index ($n_s=3.42$) in the
submillimeter region\cite{lamb} leads to a reflection at each
silicon/vacuum interface of approximately
$R=[(n_s-1)/(n_s+1)]^2\approx30\%$ per surface. This is prohibitively
large, especially for multi-lens cameras.  Nevertheless, the benefits
of silicon have motivated the development of antireflection coatings,
\cite{sherman,biber,ugras,gatesman} where the referenced AR solutions
are in IR and thus easier than mm-wave bands due to the thickness of
the AR layer. This paper details the development and testing of a
simple antireflection coating that reduces reflection to $<1.5\%$ per
lens at the design wavelength while maintaining $>90\%$ transmission
at $\nu<$ 300\,GHz.

At a fixed wavelength the ideal, normal-incidence antireflection
coating for a substrate of index $n_s$ in vacuum has an index of
refraction of $n_c=\sqrt{n_s}$ and is $t_c=\lambda_0/(4n_c)$ thick. In
our application, we are building lenses for the Atacama Cosmology
Telescope (ACT) camera\cite{act} at 145, 217, and 265\,GHz to measure
the fine scale anisotropy of the cosmic microwave background.  In the
following, we focus on the 150\,GHz band, for which the ideal
antireflection coating has $t_c=270\,\mu$m and index $n_c=1.85$.

The coating is a machined piece of
Cirlex\textsuperscript{\textregistered} \cite{fralock} polyimide glued
to silicon with Stycast\textsuperscript{\textregistered} 1266 epoxy
\cite{stycast} and Lord Ap-134 adhesion promoter\cite{lord}. For the
curved lens surface, a piece of Cirlex approximately 1\,cm thick is
machined to the curved shape and then held in a Teflon gluing jig
shaped to match the lens surface while the epoxy cures.

\section{Material Properties and Construction Details}
The low-frequency ($\sim 1$\,kHz) dielectric constant and loss
reported in the Kapton polyimide data sheet\cite{dupont} suggest
that polyimide and silicon could be combined in an AR
configuration. To ensure accurate modeling and to test sample
dependent effects, we measured the dielectric properties with
Fourier Transform Spectrometers (FTS), summarized in
Table~\ref{matprop}.

\begin{table}
 {\bf \caption{Dielectric properties of the materials
for AR coating.\label{matprop}}}
 \begin{minipage}{\textwidth}
   \centering
   \begin{tabular}{llr} \hline
 Material & Relative dielectric constant
 $\tilde{\epsilon}$\footnote{The values are fit to transmission
 measurements in the frequency range
 $100\leq\nu\leq420$\,GHz. Throughout, $\nu$ is in GHz.}\footnote{All
 results are at ambient temperature.} & $\tan\delta$\footnote{Loss
 tangent is measured at 150\,GHz.  Results have relative uncertainty
 of 10\%, or 15\% on the thinner Cirlex sample.}
  \\ \hline
Silicon ($\rho\sim5000\,\Omega\,\mathrm{cm}$) &
$11.666+i[0.0026(\nu/150)^{-1}]$ & $2.2 \times 10^{-4}$\\
Cirlex (0.25 mm) &
$3.37+i[0.037(\nu/150)^{0.52}]$ & 0.011\\
Cirlex (10 mm) &
$3.37+i[0.027(\nu/150)^{0.52}]$ & 0.008 \\
Stycast 1266   &
$2.82+i[0.065(\nu/150)^{0.27}]$ & 0.023 \\
\hline
\end{tabular}
\end{minipage}
\end{table}

Cirlex is a black pressure-formed laminate of Dupont
Kapton\textsuperscript{\textregistered} polyimide film readily
available in sheets up to 597\,mm $\times$ 597\,mm, and
thicknesses from 0.2\,mm to 3.175\,mm, with thicker constructions
possible.  In the 100--400 GHz range, at room temperature, we find
that the complex dielectric constant is well modeled as
$\tilde{\epsilon}$=$3.37+i[0.027(\nu/150)^{0.52}]$ with $\nu$ in
GHz.  At 150\,GHz the loss tangent is given by:

\begin{equation}
\tan\delta \equiv \frac{\mathrm{Im}
(\tilde{\epsilon})}{\mathrm{Re}(
\tilde{\epsilon})}=\frac{0.027}{3.37}=0.008. \label{tandel}
\end{equation}

To test the cryogenic properties of Cirlex, we placed a
250\,$\mu$m thick sample in a Bruker IFS 113 FTS that operates
between 300 and 3000\,GHz. Fitting a single complex dielectric
constant across this frequency range, we measure
$\tilde{\epsilon}=4.0+i0.06$ ($\tan\delta=0.015$) at 5\,K, and
$\tilde{\epsilon}=3.6+i0.1$ ($\tan\delta=0.030$) at room
temperature. Note that when the temperature is reduced, the loss
tangent decreases and the real part of the dielectric constant
increases.

We also performed tests using a vector network analyzer to confirm
the temperature-dependent behavior of Cirlex.  Thin slabs ($\sim$
0.02\,cm) of Cirlex were inserted into WR-10 waveguide and tested
at 90\,GHz both at room temperature and in a liquid nitrogen bath.
The full complex scattering matrix was measured over the WR-10
band (75--110\,GHz). The boundary conditions on the electric and
magnetic field at both ends of the sample lead to expressions for
the scattering parameters.  The expressions depend on the length
of the sample and its permeability and permittivity, and we invert
them to derive the complex dielectric constant.
\cite{weir,baker-jarvis} At 77\,K, Re$(\tilde\epsilon)=2.95$
($\tan\delta=0.002$) and at room temperature,
Re$(\tilde\epsilon)=3.05$ ($\tan\delta=0.017$).  Though we have
not corrected for the dimensional change of the sample upon
cooling ($\approx2\%$ shrinkage), it is clear that the loss
decreases.

Our observations of Cirlex are consistent with a smooth $20\%$
increase in Re($\tilde{\epsilon}$) from 100\,GHz to 1\,THz.  In
our calculations we use the value in Table 1, which corresponds to
a best fit of the room-temperature data over the 90--300\,GHz
range. The loss is less well constrained, though it clearly
decreases upon cooling. Table 1 shows that two similar samples
exhibited losses differing by approximately 30\%. Surface effects
may be to blame, but we also note that the difference has low
statistical significance.

Stycast 1266 is a two-component, low viscosity epoxy made by
Emerson Cuming.  We measured its properties with a cured sample
machined to be flat and 0.64\,cm thick. These properties are given
in Table~\ref{matprop}.  The index of refraction is similar to
that of Cirlex, but the loss is two to three times larger.  To
adhere the Cirlex to the silicon we coat the silicon with Lord
Ap-134 adhesion promoter before applying the Stycast.  The
adhesion promoter is needed to ensure the coating can endure
multiple cryogenic cycles.

We used high purity, high resistivity silicon for our lenses and
test samples. Recall that the complex dielectric constant for a weakly
conducting dielectric is

\begin{equation}
\tilde{\epsilon}(\nu)=\mathrm{Re}(\tilde{\epsilon})+\frac{i}{2\pi\nu\epsilon_o\rho}
\label{dielconst}
\end{equation}
in the MKS system, where Re($\tilde\epsilon$) is the real part of
the dielectric constant, $\epsilon_o=8.85 \times 10^{-14}$
($\Omega$\,cm\,Hz)$^{-1}$, and $\rho$ is the resistivity. For a
$\rho=5000\,\Omega$-cm sample at 150\,GHz, this corresponds to
$11.67+i(0.0024)$, close to the measured value.  Prior studies of
the loss in high-resistivity silicon show a complicated
temperature dependence.\cite{parshin}  However, the loss is always
smaller at 4\,K than at room temperature for potential optical
components.

Some care must be exercised when measuring the optical properties
of silicon because both heating and ultraviolet light raise its
conductivity. The nitrogen source used in the FTS was a critical
improvement over the mercury arc lamp used initially. The arc lamp
produced UV and heated the silicon samples to $70^\circ$C or
higher.  We have observed both effects affect the resistivity of
the silicon (and hence its optical transmission); heating to
$70^\circ$C alone reduced the resistivity by a factor of five.

In our experience, when a plastic sheet is glued to a glass,
quartz or silicon substrate and cooled to liquid nitrogen
temperatures, differential thermal contraction can shear apart the
substrate.  Silicon and Stycast 1266 thermally contract from
296\,K to 4\,K with $\triangle L/L=2.2 \times 10^{-4}$ and $110
\times 10^{-4}$, respectively.\cite{barron,packard} We expect
Cirlex to have a coefficient of thermal expansion comparable to
that of a plastic, {i.e.} approximately that of Stycast 1266
and ten to one hundred times that of silicon.  Through repeated
testing, we have found that the composite structure does not
fracture upon cooling if a thin layer of adhesion promoter is
applied before the epoxy.

We experimented with multiple antireflection-coated samples to test
their robustness, including five flats---four 100\,mm
in diameter and one over 200\,mm in diameter---and three
plano-convex lenses.  Both the 100\,mm and the 200\,mm flats were
dunked from room temperature into liquid nitrogen over fifty times
without damage. The lenses have been cycled four times to less
than 4\,K in a dewar, also without damage.

\section{Measurements and Modeling}

We have measured both the reflection and transmission of the AR-coated
samples.  The reflectometer is quick and straightforward to use,
though it is limited to only one frequency.  The transmission
measurements are necessary to understand the indices and the
absorption losses, and they can be made at a wide range of
frequencies.

\subsection{Reflection measurements}

\begin{figure}
  \centering
  \includegraphics[width=\textwidth]{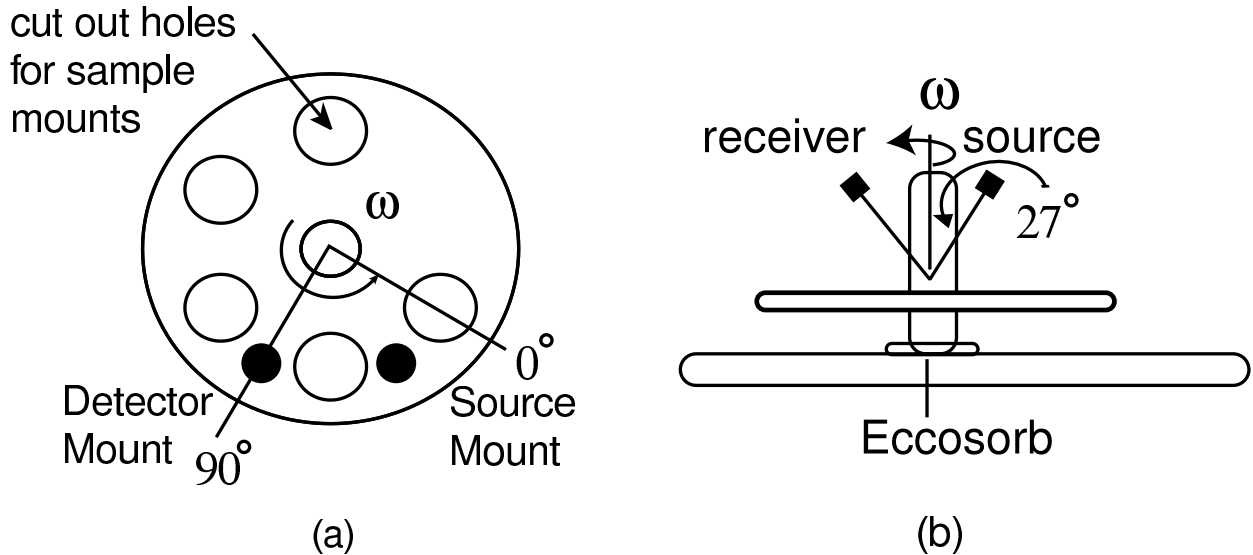}
  \caption{ \label{reflc} 
    Schematic top (a) and side view (b) of the reflectometer.  The
    axes in (a) correspond to the phase of rotation (x-axis) in
    Figure~\ref{plot}. The plate rotates via motor and belt at
    $\omega=$7.5\,rad/sec under a fixed source and receiver.  Five
    samples can be viewed in one revolution.}
\end{figure}

We built the reflectometer shown in Figure~\ref{reflc} in order to
measure the reflection of the samples. Samples are mounted in
10.16\,cm diameter holes on a 53.9\,cm diameter, 1.23\,cm thick
rotating aluminum plate. Care is taken to ensure that the sample
surface is in the same plane as the aluminum plate ($\pm
25\,\mu$m), and that the sample holder does not cause extraneous
reflections.  The diode detector, which measures total power, and
a temperature-compensated 144.00\,GHz source are mounted at a
$27^\circ$ angle from normal incidence. The feed horns for the
source and receiver are aligned so that the electric field is
oriented normal to the plane of incidence (TE mode). As the disk
rotates, the receiver successively views aluminum, a sample to be
measured, or Eccosorb through an empty sample holder. The signal
is synchronously binned and averaged over $\approx 1000$ rotations
of the plate.  The data are then scaled by taking Eccosorb to have
negligible reflection $(R=0.001)$ and aluminum to reflect
perfectly.

\begin{figure}
  \includegraphics[width=\textwidth]{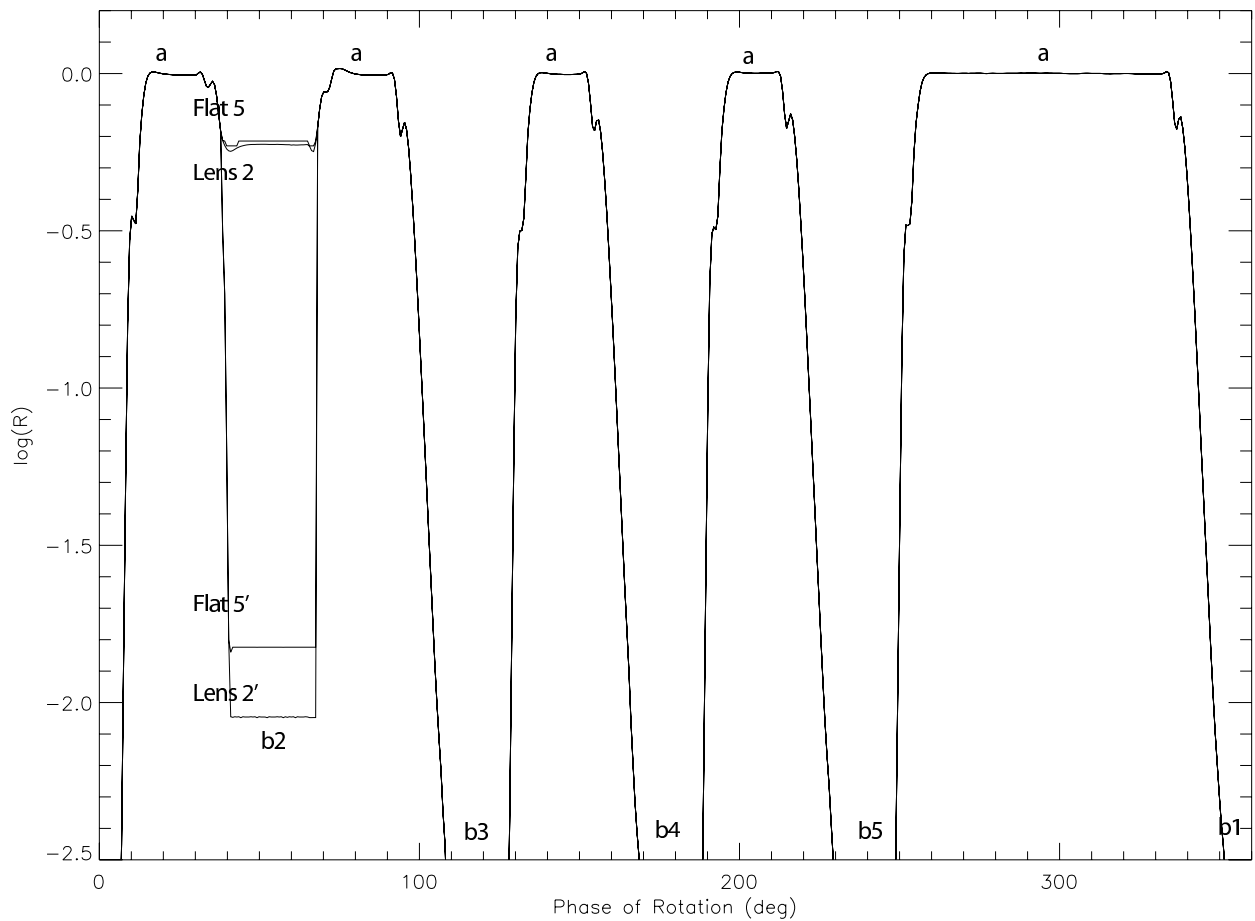}
  \caption{Four overlaid reflection plots from the reflectometer. Two
  samples are shown both before and after (denoted by primes) applying
  antireflection coating.  The reflectometer measures aluminum in
  regions labeled \emph{a}, and Eccosorb or a mounted sample at
  positions \emph{b1-b5}.  In the above plot, only \emph{b2} has a
  sample mounted. The  stability of the measurement from one sample to
  another is good, as shown by the almost perfect
  overlap of the data.}
  \label{plot}
  \end{figure}

To calibrate the reflectance $R$, we placed silicon flats of known
thicknesses (4--7\,mm) into the reflectometer.  The extremely
narrow-band light source produces interference between
reflections from the front and back of the silicon flats, so that
$R$ varies between 5 and 70\%, depending on the sample thickness.
A single model of the reflection that incorporates the incident
angle, polarization, sample thickness, frequency, and properties
of silicon fits all the data with $1\%$ mean residuals.
Figure~\ref{plot} illustrates the ease of interpreting the raw
reflectometer measurements.  All lenses are measured with the flat
side parallel to the aluminum plate.

Silicon flats and plano-convex silicon lenses were subsequently AR
coated using the procedure detailed above. The reflection was
measured before and after coating.  The results shown in
Table~\ref{befafter} demonstrate that both the flat and shaped
silicon can be AR coated to achieve as little as $1.5\%$
reflection.

\subsection{Transmission measurements}
Since we are primarily concerned with maximizing transmission
through the lenses, the transmission spectra were also measured at
room temperature on a FTS\@. The spectrometer uses the blackbody
emission of Eccosorb foam submerged in liquid nitrogen as a light
source.  The detector is a composite bolometer with a
Haller-Beeman NTD germanium thermistor and a $4\times4$\,mm
nickel-chromium coated sapphire absorber. Liquid helium cools the
bolometer to 4\,K\@.  Spectra were obtained between 100 and
425\,GHz. The small aperture of the Winston cone feeding the
bolometer sets the lower frequency limit, and a 15\,cm$^{-1}$
capacitive-inductive grid lowpass filter in the optical system
sets the upper limit. The interferometer scans continuously from
$-21$\,cm to $+21$\,cm optical path difference (OPD).  Bolometer
data are digitized at 40 samples per second by a 16-bit data
acquisition board.\cite{diamond} Acquiring each spectrum requires
eight minutes.  In analyzing the interferograms, we have discarded
data outside $\pm 18.4$\,cm OPD and apodized the rest using a
window function with Hann-like taper and a half-maximum at $\pm
13.5$\,cm optical path difference. The results are insensitive to
the details of the window, the shape of which is a compromise
between suppressing noise and not suppressing the real variations
in the transmission spectra of thick dielectric samples.

The transmission model (like the reflection model used in the
previous section) assumes that light consists of plane waves
normally incident on infinite, planar samples. The formalism is
the characteristic matrix of stratified dielectric media, given in
\S1.6 of Born and Wolf.\cite{born-wolf}  Unfortunately, the
transmission cannot readily be expressed as a function of
frequency for any system with two or more layers of dielectric.
The light incident on the samples has a focal ratio of $f/4$, slow
enough that we can ignore wavefront curvature and polarization
effects in the model. Two other effects must be considered, both
of which tend to suppress in real FTS data the extremes of the
idealized channel spectrum. First, it is necessary to treat the
model as if it had been taken on the FTS by apodizing its
``interferogram'' with the same window applied to the data. More
importantly, the model must sample the exact same frequencies as
the measurements do, where the frequency step size is the inverse
of twice the maximum optical path difference (in this measurement,
the frequencies are spaced at 0.8\, GHz). With these precautions,
we find an excellent match between the measured and modeled
transmission spectra of the coated (and uncoated) samples.

We were unable to measure the transmission of lenses directly.  It
was possible to obtain a spectrum of the transmitted light, but
this spectrum included the confounding factor of the lens's
optical power.  This optical gain factor can be computed in
principle, but it requires detailed knowledge of the complete FTS
optical system.  Worse, it is not robust against small variations
in the lens placement.  For example, one sample's spectrum
differed by a factor of two between successive tests in which the
lens was moved in the FTS beam by less than 5\,mm.  We find that
the primary value of lens transmission measurements is to check
for high transmission at the target frequency $\nu_T$ and at
$3\nu_T$.

\begin{table}[!t]
{\bf \caption{Reflection of Silicon Before and After AR
Coating}\label{befafter}}
\begin{minipage}{\textwidth}
\begin{center}
\begin{tabular}{ccccc} \hline
Sample & Thickness\footnote{Center thickness is given for the
  plano-convex lenses.  All lenses have edge thickness of $\sim
  3$\,mm.} (mm) &
$R$ uncoated & $R$ coated & Epoxy Thickness\footnote{This is the total
  epoxy thickness, not per side.} (mm)\\
\hline
Flat 5 & 5.0 & 0.59 & 0.009 & $<0.03$\\
Flat 6 & 7.0 & 0.05 &
0.05\footnote{Due to the large epoxy layer, the coating was less successful.} & $<0.2$\\
Flat 7 & 4.1 & 0.38 & 0.005 & $<0.03$\\ \hline
Lens 1 & 5.7 & 0.28 & 0.016 & $<0.07$\\
Lens 2 & 7.7 & 0.61 & 0.015 & $<0.03$\\
Lens 3 & 9.7 & 0.47 & 0.015 & $<0.06$\\ \hline
\end{tabular}
\end{center}
\end{minipage}
\end{table}

\begin{figure}
\centering
\includegraphics[width=\textwidth]{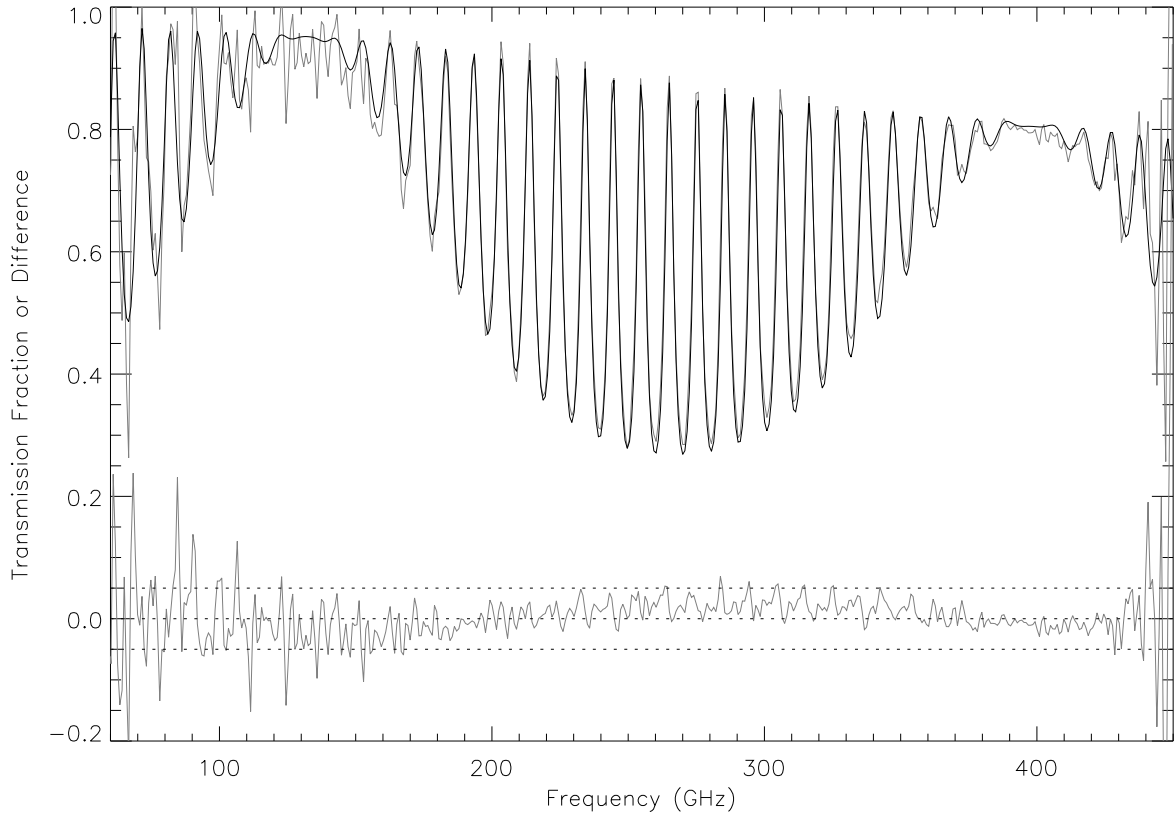}
\caption{The room temperature transmission $T$ of the coated
4\,mm-thick silicon flat (Flat
  7), both modeled (black) and measured on the FTS (gray). The measurement
  is the ratio of a sample to a reference spectrum. The lower
  curve shows that the difference (measurement minus model) is
  within 5\% of zero through the well-measured range.  The
  high transmission near 133 and 400\,GHz is due to the AR coating
  being $\lambda_0/4$ and $3\lambda_0/4$ thick.  The slow reduction in
  $T$ with increasing frequency is due to increasing loss in the
  coating and glue.  This sample was made before precise values of the
  index of Cirlex and Stycast 1266 were known.  Thus, the center of
  the passband window, 133\,GHz, is 15\,GHz below our target frequency.
} \label{data}
\end{figure}

Rather than trying to interpret the lens results, we measured the
transmission of the two coated flats labeled Flat 6 and 7 in
Table~\ref{befafter}. Figure~\ref{data} shows the transmission
spectra for one of these samples along with a model.  The
measurement is the ratio of a sample to a reference spectrum,
which are averages over two and six spectra, respectively. The
model is \emph{not} a fit to the coated transmission data but is
determined instead by the Cirlex, Stycast, and silicon properties
given in Table~\ref{matprop} and by measurements of the component
thicknesses. The one exception is the silicon loss. The coated
flats have somewhat lower resistivity (between 1300 and
$3500\,\Omega$-cm, as measured by the vendor) than the uncoated
silicon samples (all specified to exceed $5000\, \Omega$-cm); both
sets have poorly constrained $\rho$.  To handle the uncertain
silicon loss, we have treated the resistivity of the sample as an
unknown and varied it to fit the measured transmission.

\subsection{Design Considerations}
A finished design of a silicon antireflection coating for any
application requires a complete model of the system.  Incident
angle range, frequency bandwidth, and polarization  all affect the
optimal coating thickness.  It is helpful to begin the process
with a few estimates and approximate guidelines, however, and we
offer some here.  First, using Equation~\ref{dielconst} and
recalling that power loss is one $\tan\delta$ per radian of phase,
we find that absorption loss in 10\,k$\Omega$-cm silicon should
equal 1\% per centimeter, scaling as $\sigma\equiv \rho^{-1}$.

For normally incident light, the best AR-coating thickness $t_c$
can be estimated by requiring that the optical path through both
materials (coating and glue) equal one-quarter wave.  That is,
\begin{equation}
n_ct_c + n_gt_g = \lambda_0/4.
\end{equation}
Comparing this rule against the calculated reflection, we find
that it overestimates the optimal AR thickness $t_c$ by
approximately $2\,\mu$m for a glue layer $25\,\mu$m thick.  The
difference increases quadratically with glue thickness, but the
approximate expression is adequate for any reasonable size of the
glue layer.

\begin{figure}
\centering
\includegraphics[width=\textwidth]{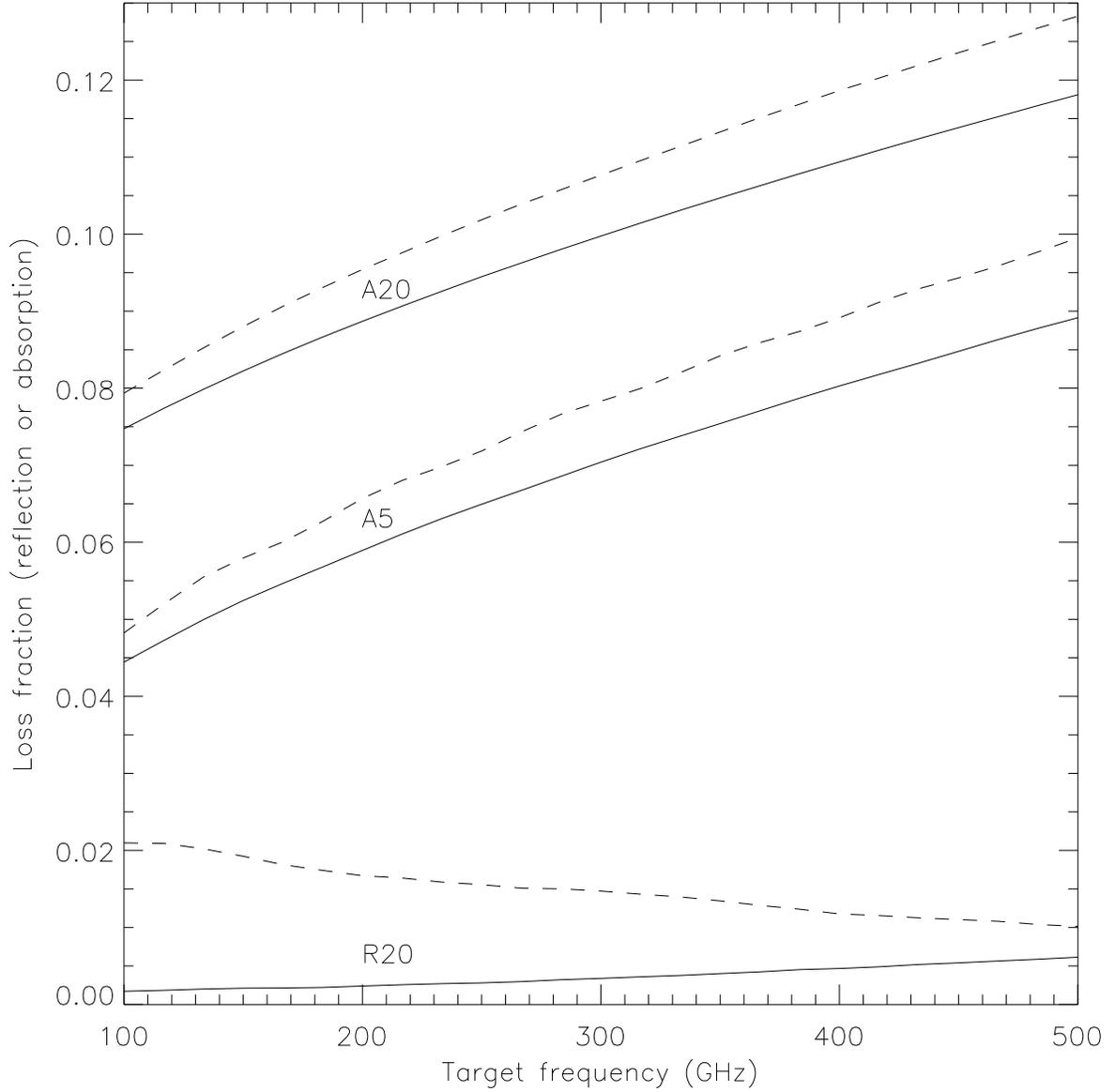}
\caption{Absorption and reflection loss modeled in two notional lenses,
5\,mm
  and 20\,mm thick.  Each lens is assumed to be coated with
  $t_g=20\,\mu$m of glue and enough Cirlex so that $(n_g t_g + n_c
  t_c)$ equals one-quarter of the vacuum wavelength.  Solid lines
  indicate normal incidence; dashed lines are for
  $40^\circ$ incident angle.  The lines labeled \emph{A5} and
  \emph{A20} show the absorption loss.  The \emph{R20} lines give the
  reflection from the 20\,mm lens; the results are not visibly
  different for reflection from the 5\,mm lens.  Absorption increases
  for oblique angles and at higher frequencies.  We
  show the expected loss at room temperature in 5000\,$\Omega$-cm
  silicon.  Cryogenically, absorption loss should be reduced.}
\label{fig.estimate}
\end{figure}

Figure~\ref{fig.estimate} shows the predicted absorption and
reflection loss calculated for a pair of realistic coated lenses,
both at $0^\circ$ and $40^\circ$ incident angles.  The model
assumes lenses made of 5000\,$\Omega$-cm silicon, 5 or 20\,mm
thick, and unpolarized light.  Over most of the frequency range it
should be possible to achieve better than 2\% reflection per lens
and less than 10\% absorption particularly if the lens is thin or
silicon resistivity is greater than the $5000\,\Omega$-cm assumed
here.

It might be possible to use the Stycast~1266 alone as an AR
coating. Its index of 1.68 is lower than the ideal 1.85, and the
loss is approximately double the loss of Cirlex.  However, cutting
a single mold to shape the curing epoxy saves three machining
steps when compared with the method for cutting Cirlex coatings
described in this article.  At 150 and 300\,GHz, the optimal
Stycast thickness are 300 and 140\,$\mu$m, which offer
transmission of 90\% and 88\%, respectively.  This compares
unfavorably with the 95\% and 93\% transmission offered by a
Cirlex-coated lens. The Stycast-coated lens would reflect
approximately 1.5\% in both frequencies, three times the
reflection expected from a Cirlex-coated lens. Still, a simpler
Stycast-only coating might suffice in some applications if the
thickness and shape could be controlled well enough.

\section{Conclusion}
We have developed and tested a technique for antireflection
coating silicon lenses at cryogenic temperatures at millimeter
wavelengths.  Flat samples show $<$ 1.5\% reflection and $>$ 92\%
transmission at the design frequency.  The remaining 6.5\% is
attributable to absorption loss, which will decrease upon cooling.

\section*{Acknowledgments}
The authors are very grateful to John Ruhl for sharing his Fourier
Transform Spectrometer and to Norm Jarosik for operating a Vector
Network Analyzer.  We thank the Princeton University Physics
Department machine shop, especially Glenn Atkinson, for developing
techniques to machine and glue the Cirlex coatings. Sarah Marriage
tested different ways of gluing silicon to polyimide;  Ted
Gudmundsen and Adrian Liu repeatedly cycled coated silicon samples
between 77 and 300\,K\@. We are grateful to our colleagues on the
ACT collaboration and to Asad Aboobaker, Andrew Bocarsly, Mark
Devlin, Simon Dicker, Phil Farese, Jeff Klein, Jeff McMahon, Amber
Miller, Mike Niemack, Suzanne Staggs, and Zachary Staniszewski for
many helpful discussions. This work was supported by the U.S.
National Science Foundation through awards AST-0408698 for the ACT
project and PHY-0355328 for the Princeton Gravity Group.


\begin{thebibliography}{99}
\bibitem{lamb}
J.W. Lamb, ``Miscellaneous data on materials for millimeter and
submillimeter optics,'' Int. J. IR and Millimeter Waves {\bf 17,}
1997--2034, (1996).

\bibitem{sherman}
G.H. Sherman and P.D. Coleman, ``Antireflection coatings for
silicon in the 2.5-50\,$\mu$m region,'' \ao {\bf 12,} 2675--2678
(1971).

\bibitem{biber}
S. Biber, J. Richter, S. Martius, and L.-P. Schmidt, ``Design of
Artificial Dielectrics for Anti-Reflection-Coatings,'' in {\it
33rd European Microwave Conference, Munich,} 1115--1118 (2003)

\bibitem{ugras}
N. G. Ugras, J. Zmuidzinas, and H. G. LeDuc, ``Quasioptical SIS
Mixer with a Silicon Lens for Submillimeter Astronomy,'' in {\it
Proceedings of the 5th International Symposium Space Terahertz
Technology,} 125.

\bibitem{gatesman}
A. J. Gatesman, J. Waldman, M. Ji, C. Musante, and S. Yngvesson,
``An Anti-Reflection Coating for Silicon Optics at Terahertz
Frequencies,'' IEEE Microwave and Guided Wave Letters, {\bf 10,}
264--266.

\bibitem{act}
J. W. Fowler, ``The Atacama Cosmology Telescope project,'' \pspie
{\bf 5498}, 1--10, (2004).

\bibitem{dupont}
E.I. du Pont de Nemours and Company, 1007 Market Street,
Wilmington, DE 19898.

\bibitem{fralock}
Fralock, Division of Lockwood Industries, Inc., 21054 Osborne
Street, Canoga Park, CA 91304.

\bibitem{stycast}
Stycast 1266, Emerson and Cuming, 869 Washington Street, Canton,
MA 02021.

\bibitem{lord}
Chemlok AP-134,  Lord Corporation, 111 Lord Drive, P.O. Box 8012,
Cary, NC 27512.

\bibitem{weir}W. B. Weir,
``Automatic measurement of complex dielectric constant and
permeability
 at microwave frequencies,'' Proc.\ IEEE, {\bf 62}, 33--36, (1974).

\bibitem{baker-jarvis}
J. Baker-Jarvis, E. J. Vanzura, \& W. A. Kissick, ``Improved
technique for determining complex permittivity with the
transmission/reflection method,'' \mtt {\bf 38}, 1096--1103,
(1990).

\bibitem{parshin}
V.V. Parshin, R. Heidinger, B.A. Andreev, A.V. Gusev, and V.B.
Shmagin, ``Silicon as an advanced window material for high power
gyrotrons,'' Int. J. IR and Millimeter Waves, {\bf 16,} 863--877,
(1995).

\bibitem{barron}
T.H.K. Barron, \& G. K. White,  ``Heat Capacity and Thermal
Expansion at Low Temperatures,'' (Kluwer Academic, New York, New
York, 1999).

\bibitem{packard}
G.W. Swift and R.E. Packard, ``Thermal contraction of Vespel
SP--22 and Stycast 1266 from 300 K to 4 K,'' Cryogenics {\bf 19,}
362--363 (1979).

\bibitem{diamond}
Diamond-MM-32-AT PC/104 board, Diamond Systems Corporation, 8430-D
Central Avenue, Newark, CA 94560.

\bibitem{born-wolf} M. Born and E. Wolf, {\it Principles of Optics,
  7th Ed.} (Cambridge, Cambridge, 1999).
\end{thebibliography}
\end{document}